\documentclass[sigconf]{acmart}

\usepackage{amsmath,epsfig}
\usepackage{pifont}
\usepackage{balance}
\usepackage{multirow}
\usepackage{booktabs}
\usepackage{cleveref}

\AtBeginDocument{%
  \providecommand\BibTeX{{%
    \normalfont B\kern-0.5em{\scshape i\kern-0.25em b}\kern-0.8em\TeX}}}

\copyrightyear{2023}
\acmYear{2023}
\setcopyright{acmlicensed}\acmConference[MM '23]{Proceedings of the 31st ACM International Conference on Multimedia}{October 29-November 3, 2023}{Ottawa, ON, Canada}
\acmBooktitle{Proceedings of the 31st ACM International Conference on Multimedia (MM '23), October 29-November 3, 2023, Ottawa, ON, Canada}
\acmPrice{15.00}
\acmDOI{10.1145/3581783.3613820}
\acmISBN{979-8-4007-0108-5/23/10}

\begin{document}

\title{BLAT: Bootstrapping Language-Audio Pre-training based on AudioSet Tag-guided Synthetic Data}

\author{Xuenan Xu}
\orcid{0000-0001-8718-1278}
\affiliation{
  \institution{Shanghai Jiao Tong University}
  \country{}
}
\email{wsntxxn@gmail.com}

\author{Zhiling Zhang}
\orcid{0000-0002-8081-704X}
\affiliation{
  \institution{Shanghai Jiao Tong University}
  \country{}
}
\email{blmoistawinde@sjtu.edu.cn}

\author{Zelin Zhou}
\orcid{0009-0002-0624-6266}
\affiliation{
  \institution{Shanghai Jiao Tong University}
  \country{}
}
\email{ze-lin@sjtu.edu.cn}

\author{Pingyue Zhang}
\orcid{0000-0002-5884-632X}
\affiliation{
  \institution{Shanghai Jiao Tong University}
  \country{}
}
\email{williamzhangsjtu@sjtu.edu.cn}

\author{Zeyu Xie}
\orcid{0009-0001-9546-3301}
\affiliation{
  \institution{Shanghai Jiao Tong University}
  \country{}
}
\email{zeyuxie29@gmail.com}

\author{Mengyue Wu}\authornote{Mengyue Wu is the corresponding author. This work has been supported by National Natural Science Foundation of China (Grant No.92048205), the Key Research and Development Program of Jiangsu Province, China (No.BE2022059-2), and Alibaba Innovative Research.}
\orcid{0000-0002-5599-8707}
\affiliation{
  \institution{Shanghai Jiao Tong University}
  \country{}
}
\email{wumengyue19@gmail.com}

\author{Kenny Q. Zhu}
\orcid{0000-0003-3782-3230}
\email{kenny.zhu@uta.edu}
\affiliation{
  \institution{University of Texas at Arlington}
  \country{}
}

\renewcommand{\shortauthors}{Xuenan Xu et al.}

\begin{abstract}

Compared with ample visual-text pre-training research, few works explore audio-text pre-training, mostly due to the lack of sufficient parallel audio-text data.
Most existing methods incorporate the visual modality as a pivot for audio-text pre-training, which inevitably induces data noise.
In this paper, we propose to utilize audio captioning to generate text directly from audio, without the aid of the visual modality so that potential noise from modality mismatch is eliminated.
Furthermore, we propose caption generation under the guidance of AudioSet tags, leading to more accurate captions.
With the above two improvements, we curate high-quality, large-scale parallel audio-text data, based on which we perform audio-text pre-training.
We comprehensively demonstrate the performance of the pre-trained model on a series of downstream audio-related tasks, including single-modality tasks like audio classification and tagging, as well as cross-modal tasks consisting of audio-text retrieval and audio-based text generation. 
Experimental results indicate that our approach achieves state-of-the-art zero-shot classification performance on most datasets, suggesting the effectiveness of our synthetic data.
The audio encoder also serves as an efficient pattern recognition model by fine-tuning it on audio-related tasks.
Synthetic data and pre-trained models are available online\footnote{The code, checkpoints and data are available at \url{https://github.com/wsntxxn/BLAT} and \url{https://zenodo.org/record/8218696}}.
\end{abstract}





\maketitle

\section{Introduction}
\label{sec:intro}

Multi-modal machine learning has become increasingly popular since it mimics our learning experience: we accept and handle information from different modalities. 
With the success of deep neural networks and large-scale datasets, we have witnessed the rapid development of multi-modal learning in recent years.
Vision-language pre-training~\cite{chen2020uniter,li2020oscar,lu2019vilbert,su2019vl} using Transformer has pushed the state of the art (SOTA) on a wide range of cross-modal tasks, such as visual question answering (VQA)~\cite{antol2015vqa}, Image-Text Retrieval~\cite{lin2014microsoft}, visual commonsense reasoning (VCR)~\cite{zellers2019recognition}, etc.
In these works, a joint representation of vision and language modalities is learned through pre-training on large-scale image-text datasets and then fine-tuned on specific downstream vision-language tasks.

\begin{figure}[!htpb]
    \centering
    \includegraphics[width=0.5\textwidth]{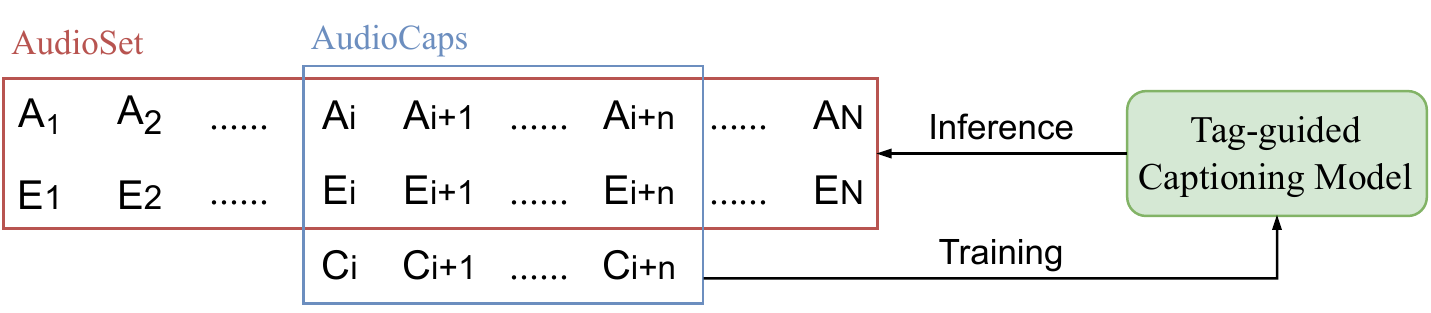}
    \caption{The illustration of the data expansion approach. ``A'', ``E'' and ``C'' denote audio, event tags and caption respectively.}
    \label{fig:bootstrapping_data}
\end{figure}

In contrast with the rapidly growing number of works in vision-language pre-training, audio-related multi-modal learning, however, is still at a preliminary stage.
Although audio is an important modality, few works explore pre-training involving audio and language.
The bottleneck of audio-language cross-modal learning lies in the scarcity of audio-text data.
Compared with large-scale image-text datasets such as COCO~\cite{lin2014microsoft} ($\sim$1.64M pairs), Visual Genome~\cite{krishna2017visual} ($\sim$5.06M pairs), and Conceptual Captions~\cite{sharma2018conceptual} ($\sim$ 12M pairs), current audio-text datasets contain only about 100K pairs (see \Cref{subsec:syn_data_generation}).
The lack of large-scale audio-text datasets may be attributed to the fact that not only the audio annotation cost is much higher than image description annotation~\cite{zhang2021enriching}, but audio-text co-occurrences are also scarcely available on the web~\cite{zhao2022connecting}.

To alleviate the above problem of data scarcity, prevailing works on audio-text cross-modal learning mostly incorporate CLIP~\cite{radford2021learning}, a powerful model enabling image-text alignment, to facilitate audio-language representation learning.
The visual modality works as a pivot to connect audio and text since video-audio co-occurrences are abundant from massive video data.
However, mismatching audio and visual modalities are commonly observed when detecting objects and events via sound and images.
For example, visible objects in videos do not necessarily make sounds while sounds may be produced by objects off the screen.
Such a mismatch leads to noise in audio-visual and audio-text alignment based on visual pivoting, indicated by the limited improvement achieved by these studies~\cite{zhao2022connecting,guzhov2022audioclip,wu2022wav2clip}.

To better circumvent the noise when expanding data, we propose an audio-captioning-based approach to expand audio-text data using AudioSet~\cite{gemmeke2017audio}, the largest free audio event dataset.
We generate captions for audio directly without the aid of the visual modality so that potential noise from modality mismatch is eliminated.
Compared with previous audio captioning works~\cite{chen2020audio,xu2021investigating}, we incorporate AudioSet tags into caption generation to improve the generated caption quality.
AudioSet contains audio clips and corresponding audio event tags in the original dataset.
Its subset AudioCaps~\cite{kim2019audiocaps} provides captions on top of tags.
Based on the provided event tags and captions, we bootstrap a tag-guided audio captioning model on AudioCaps and use it to generate large-scale audio-text data on AudioSet.
The approach is shown in \Cref{fig:bootstrapping_data}.
The bootstrapped data contains 1.22M pairs.
To this end, we propose BLAT: \textbf{B}ootstrapping \textbf{L}anguage-\textbf{A}udio pre-training based on \textbf{T}ag-guided synthetic data, where contrastive learning is used to pre-train an audio-text bi-encoder just like CLIP.

The pre-training is comprised of two phases: 1) pre-training on the large-scale synthetic data; 2) further pre-training on the real data to adapt to the real distribution.
We evaluate the performance of BLAT on a series of downstream tasks, including single-modality classification and cross-modal retrieval and generation.
Results reveal that significant achievements are achieved on all tasks by fine-tuning BLAT.
BLAT also achieves SOTA zero-shot classification performance on most datasets.

The main contribution of this paper can be summarized as follows:
\begin{itemize}
    \item We use audio captioning to curate high-quality audio-text data from audio directly, eliminating the noise from other modalities.
    \item We incorporate AudioSet tags into audio-text data generation to bootstrap large-scale synthetic data for pre-training.
    \item We validate the effect of pre-training by transferring BLAT to cross-modal and single-modality tasks, achieving significant improvements under zero-shot and fine-tuning settings.
\end{itemize}

\section{Related Work}

\subsection{Vision-Language Pre-training}
The research on multi-modal pre-trained models initially thrives in the intersection of vision and language modality. 
Vision-language pre-trained models generally handle three groups of tasks: understanding tasks like Classification, VQA and Visual Entailment, generation tasks like Image Captioning, and Image-Text Retrieval tasks.
Researchers have proposed different model structures that are specifically suitable for certain group(s) of tasks. 
Cross-encoder models process multi-modal inputs in the same encoder to allow full interaction of the two modalities and thus are generally performing well on understanding tasks \citep{chen2020uniter,li2020oscar}. 
Bi-encoder models encode the visual and textual inputs with different encoders to get separate embeddings  \citep{radford2021learning,jia2021scaling}.
Since the embeddings can be pre-computed and stored for query, they are favorable for efficient retrieval. 
Encoder-decoder models encode single or both modalities in the encoder and use a decoder for generation, which provides the capability for generation tasks \citep{wang2021simvlm,wang2022ofa}.
Our model mainly adopts the Bi-Encoder paradigm.
We exhibit that it can achieve competitive performance across all three groups of tasks.

For pre-training models, the data size has been shown to be vital for performance. 
Experimental results from the bi-encoder model CLIP show that its zero-shot image classification performance steadily increases with the number of images involved in pre-training.
Another bi-encoder ALIGN \citep{jia2021scaling} further scales up the pre-training data with noisy images from the web and shows that the models pre-trained on noisy data can still outperform those trained on higher-quality data given a larger data size.
SimVLM \citep{wang2021simvlm}, an encoder-decoder model, also achieves great success in both understanding and generation tasks with the large pre-training data ALIGN.
Inspired by their findings, we propose synthesizing parallel audio-text data for audio-language pre-training, despite the potential noise in the synthetic data.

\subsection{Audio-Language Pre-training}
With the success of visual-language pre-training, a few recent works have started to incorporate audio into multi-modal pre-training.
For instance, an audio encoder is added to CLIP with the contrastive learning paradigm.
Large-scale video-text datasets are often utilized since the dataset provides visual-text alignment while audio-visual alignment is naturally available from the video data.
VATT~\cite{akbari2021vatt} and MMV~\cite{alayrac2020self} uses HowTo100M~\cite{miech2019howto100m} and AudioSet for pre-training.
The audio-text alignment is learned implicitly through the pivot of visual modality.
AudioCLIP~\cite{guzhov2022audioclip} performs the tri-modal contrastive learning explicitly by using AudioSet event tags as the corresponding text.
Wav2CLIP~\cite{wu2022wav2clip}, in contrast, does not incorporate text into pre-training but distills CLIP by audio-visual alignment training on VGGSound~\cite{chen2020vggsound}.
Following these works, we adopt contrastive pre-training to learn audio and text representation.

Compared with either textual AudioSet tags or video descriptions, VIP$\sim$A$_\text{N}$T~\cite{zhao2022connecting} is proposed to use CLIP and the prompt ``the sound of'' to provide audio-focused descriptions for AudioSet audio clips.
A frame of the corresponding video is used as the query.
In this way, large-scale parallel audio-text pairs are automatically curated using the visual pivot.
Audio-language pre-training without explicitly incorporating the visual modality is conducted on the curated audio-text data.
Inspired by VIP$\sim$A$_\text{N}$T, we generate large-scale parallel audio-text data based on AudioSet and audio captioning.
CLAP~\cite{elizalde2023clap} is concurrent with our work.
They adopt a similar contrastive learning framework while only current parallel audio-text data is used for training.
We compare our model with these methods on zero-shot audio classification.

\subsection{Audio Event Recognition}
Audio event recognition requires recognizing the rich information in the sounds surrounding us, including the acoustic scenes where we are and what events are present.
Audio event recognition contains various tasks like acoustic scene classification~\cite{mesaros2018multi}, audio tagging~\cite{gemmeke2017audio} and sound event detection~\cite{cakir2015polyphonic}.
In recent years, the release of Detection and Classification of Acoustic Scenes and Events (DCASE) challenges encourages the development of novel datasets, tasks and approaches.
The release of AudioSet is also a milestone for audio event recognition.
It contains 2.08M 10-second audio clips\footnote{Only 1.95M clips are available in this work since some videos are removed.} with 527 annotated sound events.
Robust audio representations can be learned by pre-training a deep neural network on AudioSet.
Besides AudioSet, datasets like VGGSound and FSD50K~\cite{fonseca2022fsd50k} are also released recently to facilitate further research.

More recently, audio captioning~\cite{drossos2017automated} is proposed.
Beyond audio event tags, a caption provides an unconstrained natural language description of an audio clip.
Several datasets (see \Cref{subsec:real_datasets}) are proposed to enable audio captioning research.
Audio-text retrieval~\cite{oncescu21audio} is also proposed recently which requires retrieving audio signals using their textual descriptions and vice versa.
It should be noted that though rich textual descriptions are provided in these datasets, they are all relatively small-scale. 
The audio-language pre-training in this work is conducted based on these small-scale audio-text datasets and the large-scale audio event dataset AudioSet.
We evaluate our approach on these single-modality and multi-modal audio event recognition tasks.

\subsection{Audio Representation Learning}
Audio representation learning is an emerging field that has recently attracted increasing attention.
It involves learning general-purpose representation which can be transferred to downstream audio-related tasks.
Self-supervised speech representation~\cite{baevski2020wav2vec,hsu2021hubert,chen2022wavlm} significantly improves performance on speech-related tasks.
Audio representation~\cite{saeed2021contrastive,al2021clar,niizumi2021byol} through self-supervised learning achieves competitive results on various tasks involving speech, music and general audio.
With the release of AudioSet, many works improve the performance on audio event recognition tasks by pre-training on AudioSet~\cite{kong2020panns,gong2021audio,koutini2022efficient}.
Our work learns audio representation through audio-text contrastive learning.
We validate the effectiveness of our approach by comparing it with self-supervised COLA~\cite{saeed2021contrastive} and tag-supervised PANNs~\cite{kong2020panns}.

\section{Bootstrapping Language-Audio Data with AudioSet Tags}
In this work, we use both currently available audio-text datasets and synthetic parallel audio-text data for pre-training.
We describe these datasets and the tag-guided data generation approach in this section. 

\subsection{Current Audio-Text Datasets}
\label{subsec:real_datasets}

\begin{table}[ht]
    \centering
    \begin{tabular}{c||ccc|c|c}
    \toprule
    \multirow{2}{*}{Dataset} & \multicolumn{3}{c|}{\# Audio-text pairs} & \multirow{2}{*}{Avg \# words} & \multirow{2}{*}{Duration /h} \\
    \cline{2-4}
     & train & val & test & & \\
    \midrule
    AudioCaps & 49838 & 2475 & 4875 & 8.80 & 140\\
    \hline
    Clotho & 19195 & 5225 & 5225 & 11.33 & 37\\
    \hline
    MACS & \multicolumn{3}{c|}{17275} & 9.25 & 11\\
    \midrule
    Total & 86308 & 7700 & 10045 & 9.59 & 188\\
    \bottomrule
    \end{tabular}
    \caption{Statistics of current English audio-text datasets.}
    \label{tab:real_datasets}
\end{table}

Current parallel audio-text datasets are from audio captioning, including AudioCaps~\cite{kim2019audiocaps}, Clotho~\cite{drossos2020clotho} and MACS~\cite{martin2021diversity}.
AudioCaps is a subset of AudioSet, containing about 50K audio clips.
Each audio clip in the training set has one caption annotation while five annotations are provided for audio clips in the validation and test set.
Clotho contains 5,929 audio clips with five caption annotations provided for each clip.
The audio data are collected from Freesound~\cite{font2013freesound} platform.
MACS is a recently released dataset built on TAU Urban Acoustic Scenes 2019 dataset containing 3,930 audio clips.
Each audio clip is accompanied by several captions, ranging from two to five.
MACS does not provide splits of training, validation or test.
A summary of these datasets is in \Cref{tab:real_datasets}.

\subsection{Audio-Text Data Generation from AudioSet}
\label{subsec:syn_data_generation}

\begin{figure}[ht]
    \centering
    \includegraphics[width=0.95\linewidth]{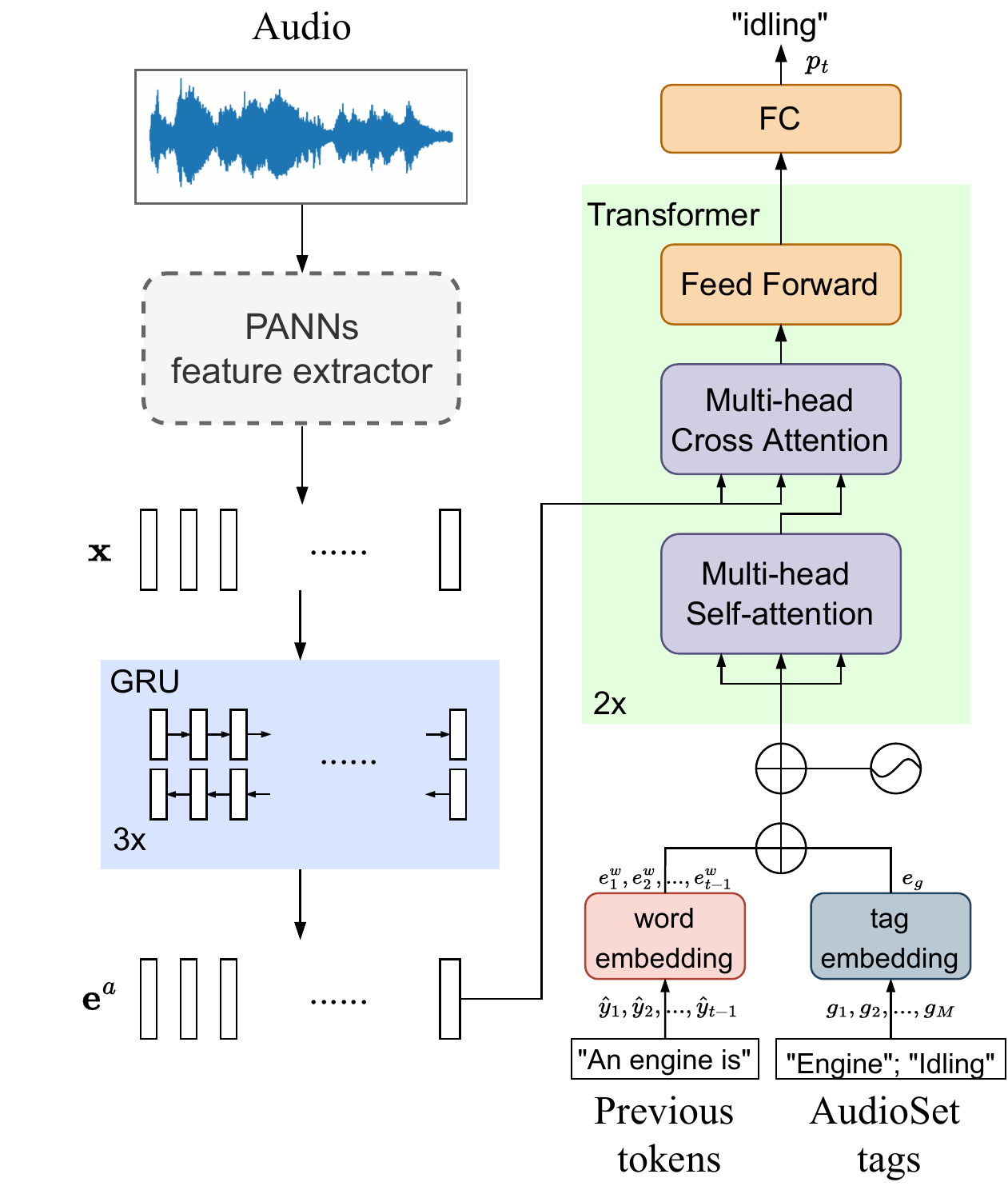}
    \caption{The proposed audio captioning system with AudioSet tag guidance. The system generates caption based on both the input audio clip and the provided AudioSet tags.}
    \label{fig:tag_caption_model}
\end{figure}

Although about 100K audio-text pairs are available in current audio-text datasets, the dataset size is much smaller than image-text datasets (e.g., $\sim$1.64M pairs in COCO, see \Cref{sec:intro}).
However, large-scale audio event data are available from AudioSet. 
To leverage the large-scale audio-only data without caption description, we aim to generate captions for audio clips in AudioSet.
Since AudioCaps is a subset of AudioSet, we first train a captioning model on AudioCaps and then use it to generate parallel audio-text data from AudioSet.
\begin{figure*}[!htpb]
    \centering
    \includegraphics[width=\linewidth]{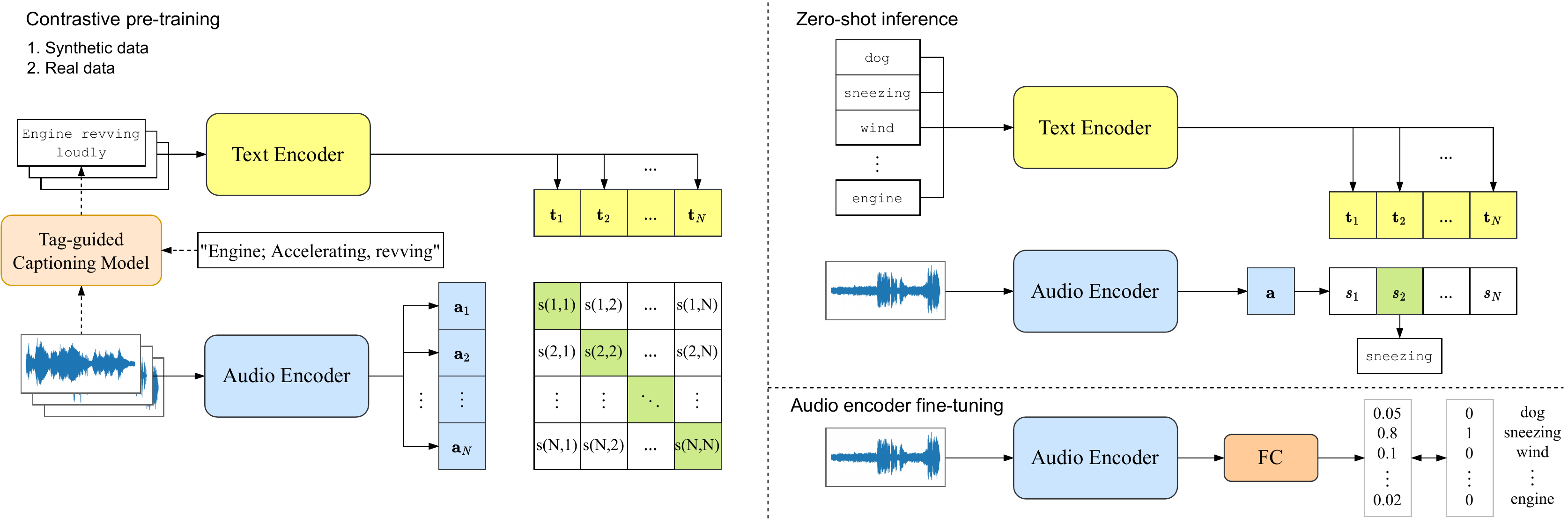}
    \caption{An overview of our proposed language-audio pre-training approach. We use a tag-guided audio captioning model to generate audio-text data. Then we conduct contrastive learning similar to CLIP (dashed lines indicate the captioning model is not involved in the pre-training) in two stages. The pre-trained model can be transferred by zero-shot inference or fine-tuning.}
    \label{fig:BLAT_framework}
\end{figure*}
Recent works tend to make use of supplementary information to guide captioning such as keyword~\cite{eren2020semantic} and similar captions~\cite{koizumi2020audio}).
However, these systems often suffer from poor prediction accuracy of supplementary information since the guidance can only be inferred from the input audio during inference.
In AudioSet, the label, consisting of audio event tags presented in the audio clip, serves as effective guidance since it is available for all clips.
Therefore, to enhance the quality of generated captions, we incorporate the event tags into caption generation.
The model generates a caption conditioned on both the input audio and the hint from AudioSet tags.
The architecture is shown in \Cref{fig:tag_caption_model}.
It contains an audio encoder and a text decoder.
A sequence of audio features $\mathbf{x}$ is fed to the encoder and transformed into a sequence of high-level representations $\mathbf{e}^{a}$.
\begin{equation}
    \mathbf{e}^{a} = \text{Encoder}(\mathbf{x})
\end{equation}
The decoder predicts the probability of each token at the time-step $t$ conditioned on partly decoded tokens $\{\hat{y}_n\}_{n=1}^{t-1}$, the provided AudioSet tags $\{{g}_m\}_{m=1}^{M}$ ($M$ is the number of tags) and $\mathbf{e}^{a}$:

\begin{align}
    p_t &= \text{Decoder}(\mathbf{e}^{a}, \{\hat{e}_n\}_{n=1}^{t-1})\\
    \hat{e}_n &= {e}_{n}^{w} + {e}^{g} \\
    {e}_{n}^{w} &= \text{WE}(\hat{y}_n), \quad {e}^{g} = \frac{1}{M} \sum_{m=1}^{M}{\text{TE}(g_m)}
\end{align}


where WE and TE denote word embedding and tag embedding layers, transforming $\hat{y}_n$ and ${g}_m$ into fixed-dimensional vectors.
Starting from the special ``$<$BOS$>$'' token, the decoder auto-regressively predicts the next token until ``$<$EOS$>$'' is reached.

In this work, we utilize deep embeddings from PANNs, specifically the \textit{CNN14} variant, as the input audio feature $\mathbf{x}$.
The encoder is a three-layer bidirectional gated recurrent unit (GRU) following \cite{eren2020semantic} while the decoder is a two-layer Transformer with the final fully connected (FC) layer.
The captioning system is trained by word-level cross entropy (CE) loss:


\begin{equation}
\mathcal{L} = \sum_{t=1}^T-\log\left(p_t(y_t)\right)
\end{equation}

where $y_t$ is the ground truth token at the time-step $t$.

After training the AudioSet tag-guided captioning model, we use it to generate captions for large-scale AudioSet audio clips.
However, the data distribution of AudioCaps is different from AudioSet since audio clips with specific event tags are excluded during the construction process of AudioCaps~\cite{kim2019audiocaps}.
To circumvent the distribution bias problem, we exclude audio clips with tags that never appear in AudioCaps, with about 1.22M audio clips left.
One caption is generated for each audio clip using the enhanced captioning model, resulting in about 1.22M audio-text pairs.
We use this large-scale synthetic parallel audio-text data for pre-training.

\section{Audio-Text Pre-training}
In this section, we describe the proposed framework.
The framework consists of an audio encoder and a text encoder for the two modalities.
As \Cref{fig:BLAT_framework} shows, the model is first pre-trained by contrastive learning.
The pre-training consists of two steps: 1) pre-training on synthetic parallel audio-text data; 2) further pre-training on the real data.
Since there is a gap between the quality of real and synthetic data, the second pre-training step is adopted to alleviate the bias caused by synthetic data.
We use the combination of all training sets of real audio-text data introduced in \Cref{subsec:real_datasets} for training.

After pre-training, the pre-trained model is transferred to several kinds of downstream tasks.
Take audio classification as an example, the pre-trained model can be used for both zero-shot inference and fine-tuning.
For zero-shot inference, the similarity scores between the audio clip and all textual labels are calculated as the estimated probabilities.
For fine-tuning, a fully-connected (FC) layer is appended after the audio encoder for further classification fine-tuning to boost performance.

We first illustrate the contrastive pre-training approach.
Then the architectures of the two encoders are introduced respectively.

\subsection{Contrastive Pre-training}
Similar to CLIP, the proposed contrastive learning approach learns the correspondence between the text content and the audio events in an arbitrary audio-text pair.
For an audio clip $\mathcal{A}$ and a sentence $\mathcal{T}$, the audio and text encoders $\text{Enc}_A$ and $\text{Enc}_T$ transform them into two embeddings $\mathbf{a}$ and $\mathbf{t}$ respectively.
A multi-modal embedding space is learned by maximizing the similarity between $\mathbf{a}$ and $\mathbf{t}$ of matched audio-text pairs and minimizing that of mismatched pairs.
Following CLIP, the training objective is to minimize the InfoNCE loss~\cite{van2018representation}.
Given a minibatch of $N$ audio-text pairs $(\mathcal{A}_1, \mathcal{T}_1), (\mathcal{A}_2, \mathcal{T}_2),\ldots, (\mathcal{A}_N, \mathcal{T}_N)$, their embeddings are calculated:

\begin{align}
\begin{split}
    \mathbf{a}_i = \text{Enc}_A (\mathcal{A}_i)\\
    \mathbf{t}_i = \text{Enc}_T (\mathcal{T}_i)
\end{split}
\end{align}
The training loss is a symmetric cross entropy loss between the predicted cosine similarity scores and the ground truth pairing labels:
\begin{align}
    \begin{split}
        &s(i, j) = \frac{\mathbf{a}_i\cdot\mathbf{t}_j^{\mathrm{T}}}{
        \Vert \mathbf{a}_i \Vert \cdot \Vert \mathbf{t}_j \Vert
        }\\
        &\mathcal{L}_i^{A \rightarrow T} = -\log \frac{exp\left(s(i, i\right) / \tau )}{\sum_{j=1}^N exp(s\left(i, j\right) / \tau)}\\
        &\mathcal{L}_i^{T \rightarrow A} = -\log \frac{exp\left(s(i, i\right) / \tau )}{\sum_{j=1}^N exp(s\left(j, i\right) / \tau)}\\
        &\mathcal{L} = \frac{1}{N}\sum_{i=1}^N(\mathcal{L}_i^{A \rightarrow T} + \mathcal{L}_i^{T \rightarrow A})
    \end{split}
\end{align}
where $\tau$ is the temperature optimized jointly with $\text{Enc}_A$ and $\text{Enc}_T$.

\subsection{Audio Encoder}
Similar to the feature extractor in \Cref{subsec:syn_data_generation}, we use the pre-trained CNN14 from PANNs~\citep{kong2020panns} as $\text{Enc}_A$ instead of training the model from scratch.
Time-frequency representation Log Mel Spectrogram (LMS) is extracted from the input audio and fed to 12 convolution blocks.
$2\times2$ max pooling is done between every two blocks.
After the convolution blocks, the audio embedding \textbf{a} is obtained by a global pooling on the feature map and transformation through a fully-connected layer. 
Although Transformer-based models are applied for audio classification recently~\cite{gong2021audio} and achieve better performance than convolutional neural networks (CNN), it works on a sequence of patch embeddings without sub-sampling, resulting in high memory demand.
Therefore, we adopt the pre-trained CNN14 to enable a larger minibatch size for training.

\subsection{Text Encoder}
For the text encoding part, we utilize BERT to transform $\mathcal{T}$ into $\mathbf{t}$.
It is a deep Transformer pre-trained on large-scale corpora, including BooksCorpus and English Wikipedia, by self-supervised learning.
Due to its powerful capability to extract representations with contextual semantics, BERT has exhibited superior performance on a series of language understanding tasks~\cite{devlin2019bert}.
In this work, we employ BERT$_\text{MEDIUM}$~\cite{turc2019well} as Enc$_T$ for better computation efficiency and lower memory requirements.
It consists of eight Transformer layers with a hidden embedding size of 512.

\section{Experimental Setup}
In this section, we present our experimental setup for expanding audio-text data, pre-training, fine-tuning and downstream evaluation.

\subsection{Synthetic Audio-Text Data Generation}
\label{subsec:syn_data_generation_setup}

The AudioSet tag-guided captioning model takes the feature extracted by PANNs CNN14 as the input.
The encoder is a 3-layer bidirectional GRU and the decoder is a 2-layer Transformer.
Details can be referred to \cite{xu2022sjtu}.
The model is trained on AudioCaps for 25 epochs with a batch size of 64.
The learning rate warms up to $5\times10^{-4}$ and then exponentially decays to $5\times10^{-7}$ until the end.
We use beam search with a size of 3 when expanding audio-text pairs.

\subsection{Pre-training}

In the first pre-training step, we use a batch size of 128 and train the model for 200K iterations.
About 1,200 audio-text pairs are randomly selected from the synthetic data to form a separate validation set.
The model is validated every 500 iterations on the validation set.
We use the Adam optimizer with the maximum learning rate of $1\times10^{-4}$.
The learning rate is decayed by a cosine scheduler~\cite{loshchilov2016sgdr} with linear warm up in the first 10k iterations.

The model with the best performance on the synthetic validation set is used to initialize parameters for the second pre-training step.
The setup is similar to the first step with several modifications on hyper-parameters.
The total training iterations and warm up iterations are 15000 and 750 while the model is validated every 750 iterations.

\subsection{Downstream Evaluation}

\begin{table}[ht]
    \centering
    \begin{tabular}{cccc}
    \toprule
    Task & Dataset & \# Audio clips & Metric \\
    \midrule
    Audio-text & AudioCaps & 50K & \multirow{2}{*}{R@K} \\
    Retrieval & Clotho & 6K & \\
    \hline
    \multirow{2}{*}{\shortstack{Audio\\Captioning}} & AudioCaps & 50K & COCO \&  \\
     & Clotho & 6K & FENSE \\
    \hline
    \multirow{3}{*}{Classification} & ESC50 & 2K & \multirow{3}{*}{Accuracy}\\
     & US8K & 8K & \\
     & VGGSound & 192K & \\
    \hline
    \multirow{2}{*}{Tagging} & FSD50K & 50K & \multirow{2}{*}{mAP}\\
     & AudioSet & 1.93M & \\
    \bottomrule
    \end{tabular}
    \caption{A summary of downstream cross-modal and single-modality tasks. US8K is the abbreviation of UrbanSound8K.}
    \label{tab:task_summary}
\end{table}

The pre-trained BLAT can be transferred to a series of downstream tasks, which is summarized in \Cref{tab:task_summary}, including both cross-modal tasks and single-modality tasks.
\textbf{Cross-modal audio-text tasks} include \textit{\textbf{audio-text retrieval}} and \textit{\textbf{audio captioning}}.
For audio-text retrieval, we use recall at K (R@K) as the evaluation metric.
Standard COCO evaluation metrics from image captioning are used to evaluate audio captioning performance.
Besides, we also incorporate FENSE~\cite{zhou2022can} into evaluation for its higher correlation with human judgments.

\textbf{Single-modality tasks} include \textit{\textbf{single-label (classification)}} and \textit{\textbf{multi-label (tagging) audio classification}}.
Accuracy and mean average precision (mAP) are used for evaluation.
We include several datasets with the size ranging from 2K to 1.93M for comparison with previous works.

\subsection{Zero-shot Classification}
With the pre-trained BLAT, we can perform zero-shot classification.
If a textual label contains ``\_'', we replace ``\_'' with a blank.
BLAT calculates the similarity scores between a given audio clip and all these textual labels.
These scores are treated as the predicted probability of each audio event for evaluation.

\subsection{Fine-tuning}
Fine-tuning is commonly adopted to transfer the general-purpose pre-trained model to downstream tasks that may focus on specific domains.
We illustrate the fine-tuning procedures for two cross-modal tasks and single-modality audio classification respectively.  
\subsubsection{Audio-text Retrieval}
The fine-tuning on audio-text retrieval tasks uses almost the same configuration as the pre-training step.
For both AudioCaps and Clotho, we fine-tune the pre-trained bi-encoder model for 20 epochs using the InfoNCE loss with a batch size of 128.
The learning rate linearly warms up to the maximum value in the first epoch.
The maximum learning rate for AudioCaps and Clotho is $5\times10^{-5}$ and $2\times10^{-6}$, respectively.

\subsubsection{Audio Captioning}
The audio captioning system is similar to the model in \Cref{subsec:syn_data_generation} except 1) the audio feature is extracted by BLAT instead of PANNs; 2) the system does not receive guidance from AudioSet tags.
For both AudioCaps and Clotho, the training and inference configuration follows \Cref{subsec:syn_data_generation_setup}.

\subsubsection{Audio Classification and Tagging}
For single-modality tasks, we further fine-tune the pre-trained audio encoder $\text{Enc}_A$.
An extra FC layer is added to $\text{Enc}_A$ for classification.
We perform two types of fine-tuning: linear probing and fine-tuning the whole $\text{Enc}_A$.
For linear probing, $\text{Enc}_A$ is used as a feature extractor and only the final FC layer is trained while no parameters are frozen in the second setting.
Cross entropy loss and binary cross entropy loss are used for classification and tagging training respectively.

\section{Results}

In this section, we present the comprehensive performance of BLAT.
We first evaluate the quality of bootstrapped synthetic audio-text data.
Then we reveal the influence of pre-training on downstream tasks.
In experiments where only the audio encoder is used, we take PANNs~\cite{kong2020panns} for comparison since both models share the same CNN14 architecture and use AudioSet for pre-training.
We also incorporate self-supervised audio representation COLA~\cite{saeed2021contrastive}.
For all experiments except pre-training, we report results based on three randomly seeded runs.

\subsection{Benefits of Bootstrapped Audio-Text Data}

\begin{table}[ht]
    
    \begin{tabular}{c|cccccc}
    \toprule
    & $\text{B}_4$ & R & M & C & S & F\\
    \midrule
    Synthetic w/o tag & 24.1 & 47.0 & 23.1 & 71.2 & 19.2 & 60.1\\
    Synthetic & 26.4 & 49.0 & 24.5 & 80.4 & 21.0 & 62.5\\
    Human & 29.0 & 49.5 & 28.8 & 90.8 & 28.8 & 68.0\\
    \bottomrule
    \end{tabular}
    \caption{The comparison of synthetic parallel audio-text data and real data in terms of audio captioning performance. Metrics include $\text{BLEU}_4$ ($\text{B}_4$), $\text{ROUGE}_\text{L}$ (R), METEOR (M), CIDEr (C), SPICE (S) and FENSE (F).}
    \label{tab:syn_data_quality}
\end{table}

\begin{table*}[!htpb]
    \centering
    \begin{tabular}{c|c||cccc||cccc}
    \toprule
    \multirow{3}{*}{\shortstack{Training \\On Target}} &
    \multirow{3}{*}{Configuration} & \multicolumn{4}{c||}{AudioCaps} & \multicolumn{4}{c}{Clotho} \\
    \cline{3-10}
    & & \multicolumn{2}{c}{Audio $\Rightarrow$ Text} & \multicolumn{2}{c||}{Text $\Rightarrow$ Audio} & \multicolumn{2}{c}{Audio $\Rightarrow$ Text} & \multicolumn{2}{c}{Text $\Rightarrow$ Audio} \\
    & & R@1 & R@10 & R@1 & R@10 & R@1 & R@10 & R@1 & R@10 \\
    \midrule
    \multirow{3}{*}{No} & VIP$\sim$A$_\text{N}$T~\cite{zhao2022connecting} & 15.2 & 52.9 & 9.9 & 45.6 & 7.1 & 30.7 & \textbf{6.7} & \textbf{29.1} \\
    & template tags & 12.7 & 49.8 & 8.9 & 43.2 & 6.0 & 24.9 & 4.3 & 23.0 \\
    & BLAT & \textbf{32.6} & \textbf{76.7} & \textbf{23.5} & \textbf{68.4} & \textbf{7.6} & \textbf{31.5} & 5.6 & 23.8 \\
    \midrule
    \multirow{2}{*}{Yes} & from scratch & $40.4_{\pm 1.5}$ & $85.7_{\pm 0.7}$ & $33.3_{\pm 0.3}$ & $82.4_{\pm 0.3}$ & $13.9_{\pm 0.2}$ & $48.2_{\pm 1.4}$ & $12.3_{\pm 0.6}$ & $46.1_{\pm 0.9}$\\
    & BLAT fine-tuning & $47.5_{\pm 0.4}$ & $87.6_{\pm 0.2}$ & $38.2_{\pm 0.1}$ & $85.1_{\pm 0.1}$ & $17.9_{\pm 0.4}$ & $50.9_{\pm 1.9}$ & $13.7_{\pm 0.4}$ & $48.9_{\pm 0.5}$\\
    \bottomrule
    \end{tabular}
    \caption{Audio-text retrieval performance. The upper half denotes pre-training on different synthetic data and evaluating the pre-trained model without fine-tuning on the target dataset. The lower half shows the performance of training our model on the target dataset. R@K denotes recall at K.}
    \label{tab:pre_train_effects}
\end{table*}

\begin{table*}[ht]
    \centering
    \begin{tabular}{c|c||cccccc}
    \toprule
    Dataset & Audio Feature & $\text{B}_4$ & R & M & C & S & F\\
    \midrule
    \multirow{3}{*}{AudioCaps} & COLA & $14.1_{\pm 0.2}$ & $36.6_{\pm 0.4}$ & $15.7_{\pm 0.2}$ & $30.7_{\pm 0.8}$ & $10.0_{\pm 0.0}$ & $38.6_{\pm 0.3}$
    \\
    & PANNs & $\mathbf{27.3_{\pm 0.4}}$ & $\mathbf{49.7_{\pm 0.2}}$ & $24.4_{\pm 0.1}$ & $72.3_{\pm 0.8}$ & $18.1_{\pm 0.2}$ & $60.6_{\pm 0.4}$ \\
     & BLAT & $27.2_{\pm 0.2}$ & $49.5_{\pm 0.0}$ & $\mathbf{24.7_{\pm 0.1}}$ & $\mathbf{73.3_{\pm 0.4}}$ & $\mathbf{18.4_{\pm 0.2}}$ & $\mathbf{61.5_{\pm 0.3}}$ \\
    \midrule
    \multirow{3}{*}{Clotho} & COLA & $10.0_{\pm 0.6}$ & $31.0_{\pm 0.2}$ & $13.0_{\pm 0.2}$ & $18.7_{\pm 1.9}$ & $7.5_{\pm 0.2}$ & $29.9_{\pm 0.7}$\\
    & PANNs & $15.8_{\pm 0.3}$ & $\mathbf{37.6_{\pm 0.2}}$ & $17.5_{\pm 0.1}$ & $39.3_{\pm 0.8}$ & $12.1_{\pm 0.1}$ & $43.8_{\pm 0.4}$\\
     & BLAT & $\mathbf{16.0_{\pm 0.4}}$ & $\mathbf{37.6_{\pm 0.0}}$ & $\mathbf{17.8_{\pm 0.0}}$ & $\mathbf{41.5_{\pm 0.6}}$ & $\mathbf{12.6_{\pm 0.1}}$ & $\mathbf{45.8_{\pm 0.6}}$\\
    \bottomrule
    \end{tabular}
     \caption{A comparison of audio captioning performance using different audio features.}
    \label{tab:caption_finetune}
\end{table*}

\subsubsection{Data Quality Comparison on Captioning}

The quality of bootstrapped data is first evaluated in terms of captioning performance.
We compare the performance of synthetic captions and human-annotated captions on AudioCaps test set.
Since human annotations are used both as the candidate to be evaluated and the reference, we use a round-robin evaluation schedule.
Specifically, we exclude one reference annotation in each round and evaluate the caption based on the left four annotations.
The five scores are averaged as the performance indicator.
We compare the performance of synthetic and human-annotated captions on AudioCaps in \Cref{tab:syn_data_quality}.
We also include the captioning system without AudioSet tag guidance to show the effect of importing audio event tags.
Metrics reveal that the tag guidance brings significant improvement.
For $\text{ROUGE}_\text{L}$, the synthetic data performance is surprisingly comparable with human annotation.
In terms of metrics evaluating the semantic similarity like SPICE and FENSE, human annotation is still much better.
The model is capable of generating high-quality captions with the tag guidance though there is still a quality gap between the synthetic and real data.

\subsubsection{Data Quality Comparison on Retrieval}
We also conduct the bootstrapped data in terms of zero-shot audio-text retrieval performance, shown in the upper half of \Cref{tab:pre_train_effects}.
We compare our data with VIP$\sim$A$_\text{N}$T~\cite{zhao2022connecting}, which uses CLIP and the prompt ``the sound of'' to retrieve captions from AudioCaps and Clotho training corpus.
The two synthetic datasets share a similar size (1.22M and 1.08M).
For comparison, we also use a simple template ``The sound of \textless tag 1 \textgreater, \textless tag 2 \textgreater, ..., and \textless tag n \textgreater'' to convert AudioSet tags into captions, denoted as ``template tags'' in the table.
BLAT significantly outperforms template tags on two datasets, indicating that our tag-guided captioning model can generate text data of higher quality.
This is likely attributed to missing annotations in AudioSet (as elaborated in \Cref{subsubsec:zero_shot_transfer}): missing tags make the template-based text less specific (e.g., the sound of speech) and comprehensive than that generated by our captioning model (e.g., a woman is speaking while something is being fried) bootstrapped from AudioCaps.
The comparison between BLAT and VIP$\sim$A$_\text{N}$T shows that the model trained on our synthetic data significantly outperforms VIP$\sim$A$_\text{N}$T except for text-to-audio retrieval on Clotho.
It indicates that using the visual modality as a pivot between audio and text leads to noisy data.
The noise may come from the asynchronous audio and visual modalities.
Note that Clotho captions are used to curate audio-text data in VIP$\sim$A$_\text{N}$T while in our work only AudioCaps is used.
The distribution difference between AudioCaps and Clotho captions~\cite{martin2021diversity} leads to our model's unsatisfactory performance on Clotho.

\subsection{Cross-modal Audio-and-Language Tasks}

The lower half of \Cref{tab:pre_train_effects} shows the performance of transferring BLAT to audio-text retrieval.
We compare the model fine-tuning from BLAT with one trained from scratch.
As the size of Clotho is small, the model trained from scratch performs poorly.
With the initialization from BLAT, significant improvement can be witnessed on both AudioCaps and Clotho.

The performance of BLAT transferred to audio captioning is shown in \Cref{tab:caption_finetune}.
Without the supervision of event labels or textual descriptions, self-supervised COLA performs much worse than PANNs and BLAT.
BLAT feature outperforms PANNs mainly on metrics regarding the semantic content like CIDEr and FENSE.
This indicates that BLAT feature is more representative and helps the model generate more relevant descriptions.

\subsection{Single-modality Audio Classification}

\begin{table*}[ht]
    \centering
    \begin{tabular}{cc|c||ccccc}
    \toprule
     & Model & \# params / M & ESC50 & US8K & VGGSound (mAP) & FSD50K & AudioSet\\
    \midrule
     & SOTA & - & 97.2~\cite{guzhov2022audioclip} & 90.1~\cite{guzhov2022audioclip} & 52.5~\cite{kazakos2021slow} & 65.3~\cite{koutini2022efficient} & 47.1~\cite{koutini2022efficient}\\
    \midrule
    \multirow{5}{*}{Zero-shot} & AudioCLIP & 95.9 & 69.4 & 68.8 & - & - & -\\
     & Wav2CLIP & \textbf{74.8} & 41.4 & 40.4 & - (10.0) & 3.0 & - \\
     & VIP$\sim$A$_\text{N}$T & 151.3 & 69.2 & 71.7 & - & - & \textbf{13.3}\\
     & CLAP & 192.1 & \textbf{82.6} & 73.2 & - & 30.2 & 5.8 \\
     & BLAT & 123.7 & 80.6 & \textbf{77.3} & 14.9 (\textbf{13.5}) & \textbf{31.3} & 10.5\\
    \midrule
    \multirow{3}{*}{Linear probing} & COLA & 79.7 & $38.2_{\pm 0.3}$ & $53.8_{\pm 0.3}$ & $13.9_{\pm 0.1}$ & $10.7_{\pm 0.0}$ & $2.1_{\pm 0.1}$ \\
    & PANNs & 79.7& $89.9_{\pm 0.1}$ & $82.6_{\pm 0.3}$ & $41.4_{\pm 0.8}$ & $29.7_{\pm 0.2}$ & - \\
     & BLAT & 79.7& $\mathbf{94.8_{\pm 0.3}}$ & $\mathbf{85.7_{\pm 0.3}}$ & $\mathbf{42.9_{\pm 0.5}}$ & $\mathbf{32.4_{\pm 0.7}}$ & $\mathbf{38.7_{\pm 0.0}}$\\
    \midrule
    \multirow{3}{*}{Fine-tuning} & COLA & 79.7& $78.8_{\pm 0.7}$ & $75.3_{\pm 0.4}$ & $48.7_{\pm 0.8}$ & $47.9_{\pm 0.9}$ & $43.6_{\pm 0.2}$ \\
    & PANNs & 79.7& $95.4_{\pm 0.1}$ & $87.4_{\pm 0.2}$ & $\mathbf{55.3_{\pm 0.8}}$ & $57.6_{\pm 0.2}$ & - \\
     & BLAT & 79.7 & $\mathbf{95.8_{\pm 0.2}}$ & $\mathbf{89.0_{\pm 0.1}}$ & $54.8_{\pm 0.1}$ & $\mathbf{60.3_{\pm 0.5}}$ & $\mathbf{44.0_{\pm 0.2}}$\\
    \bottomrule
    \end{tabular}
    \caption{Audio classification and tagging performance in different settings: 1) zero-shot transfer 2) linear probing 3) fine-tuning. On VGGSound, we list mAP in parentheses to compare with Wav2CLIP.
    We only include parameters necessary for zero-shot inference when counting parameter numbers (i.e., the visual encoding part is excluded for AudioCLIP and VIP$\sim$A$_\text{N}$T).}
    \label{tab:classify_performance}
\end{table*}

\begin{figure*}[ht]
    \centering
    \includegraphics[width=0.95\linewidth]{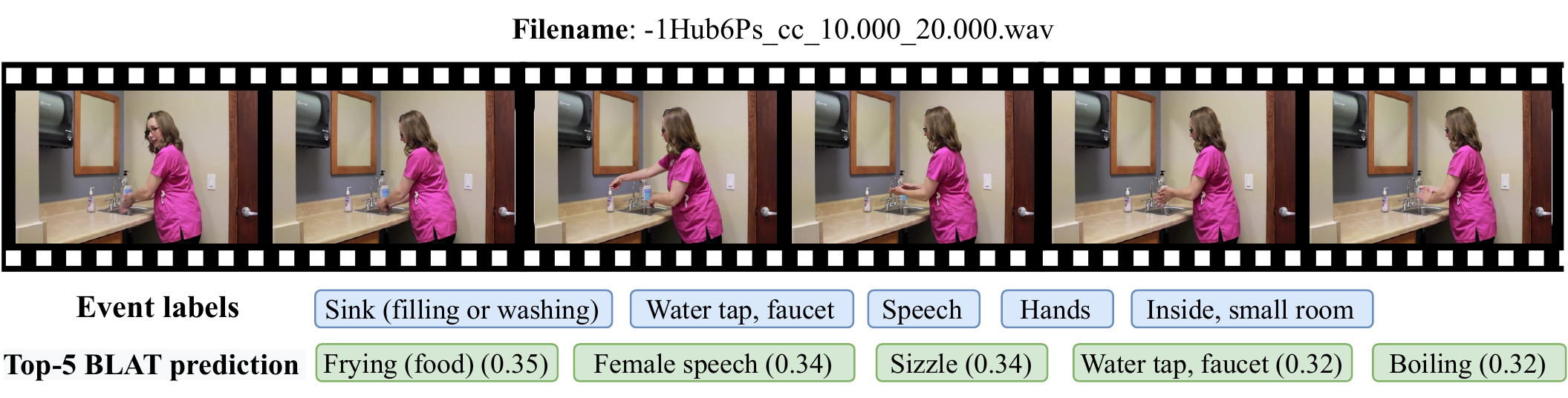}
    \caption{An example of annotation errors in AudioSet. A woman is speaking in the audio clip while the corresponding event ``Female speech, woman speaking'' is not annotated.}
    \label{fig:audioset_missing_label}
\end{figure*}

\subsubsection{Zero-shot Transfer}
\label{subsubsec:zero_shot_transfer}
Under the zero-shot setting, the transferring ability of BLAT is evaluated.
Previous works enabling zero-shot inference, including AudioCLIP~\cite{guzhov2022audioclip}, Wav2CLIP~\cite{wu2022wav2clip}, VIP$\sim$A$_\text{N}$T~\cite{zhao2022connecting}, and CLAP~\cite{elizalde2023clap} are incorporated for comparison.
Except for CLAP, CLIP is utilized for synthetic data generation or pre-training.
We also include current SOTA results as a topline for reference.
Results are shown in the upper half of \Cref{tab:classify_performance}.
The parameter numbers of these models are listed.
Compared with works relying on CLIP, BLAT achieves SOTA zero-shot performance with a moderate model size, validating the benefit of eliminating noise from the visual modality.
On VGGSound, BLAT outperforms Wav2CLIP even though the latter is pre-trained on VGGSound, indicating the transferring ability of BLAT.
However, BLAT achieves a low mAP on AudioSet.
Apart from the data distribution bias caused by the creation of AudioCaps, we observe that the noise in AudioSet labels exacerbates the problem.
Previous works reveal that AudioSet annotations often contain only part of all events presented in a clip~\cite{gong2021psla}.
An example is shown in \Cref{fig:audioset_missing_label}.
Speech from a woman can be clearly heard in the audio clip and BLAT assigns a high probability to the event ``Female speech, woman speaking''.
However, the event does not occur in the AudioSet annotation.
We assume such annotation noise leads to unreliable results.
On FSD50K where annotations are more accurate, BLAT achieves a much higher mAP.
Using a similar audio-text pre-training paradigm on real datasets, CLAP achieves similar results on ESC50 and FSD50K while BLAT outperforms CLAP on US8K and AudioSet with fewer parameters.
This validates the benefit of incorporating our bootstrapped data into pre-training.

\subsubsection{Linear probing and Fine-tuning}
The lower half of \Cref{tab:classify_performance} shows the results of transferring BLAT to audio classification by linear probing and fine-tuning.
Since PANNs are trained on AudioSet with event labels, we do not further fine-tune PANNs on AudioSet.
Like audio captioning, COLA performs much worse than PANNs and BLAT, especially under the linear probing setting.
Although COLA can be applied to any audio data, its representation does not generalize well to audio classification tasks without the supervision of labels.
In both linear probing and fine-tuning settings, BLAT outperforms PANNs on most datasets\footnote{Except VGGSound, results are significant at a level of 0.05.}.
With only one FC classifier, the performance of linear probing BLAT on ESC50 and US8K is even close to current SOTA results, indicating that BLAT serves as a powerful feature extractor.
Especially for small datasets, BLAT is able to extract highly discriminative features for classification.
By fine-tuning BLAT, we achieve results close to SOTA, which validates its transferring ability to other tasks.
The supervision of natural language exhibits a better ability to be transferred to a variety of audio classification tasks than event labels.
Note that we do not adopt training techniques like data augmentation or task-specific loss functions.

\section{Conclusion}

In this work, we propose an AudioSet tag-guided audio captioning model to bootstrap large-scale audio-text data.
Different from previous methods, the data generation approach does not incorporate video to eliminate the noise induced by the visual modality. 
Based on the bootstrapped data, we pre-train an audio-text bi-encoder using contrastive learning.
After pre-training the model on the synthetic data and the real data successively, we obtain BLAT which can be transferred to a series of downstream tasks.
Experimental results on both cross-modal and single-modality tasks, including retrieval, generation and classification, validate the effectiveness of BLAT.
Under the stringent zero-shot condition where no training data is available, BLAT exhibits SOTA performance on most datasets.

\bibliographystyle{ACM-Reference-Format}
\balance
\bibliography{refs}


\begin{thebibliography}{53}


\ifx \showCODEN    \undefined \def \showCODEN     #1{\unskip}     \fi
\ifx \showDOI      \undefined \def \showDOI       #1{#1}\fi
\ifx \showISBNx    \undefined \def \showISBNx     #1{\unskip}     \fi
\ifx \showISBNxiii \undefined \def \showISBNxiii  #1{\unskip}     \fi
\ifx \showISSN     \undefined \def \showISSN      #1{\unskip}     \fi
\ifx \showLCCN     \undefined \def \showLCCN      #1{\unskip}     \fi
\ifx \shownote     \undefined \def \shownote      #1{#1}          \fi
\ifx \showarticletitle \undefined \def \showarticletitle #1{#1}   \fi
\ifx \showURL      \undefined \def \showURL       {\relax}        \fi
\providecommand\bibfield[2]{#2}
\providecommand\bibinfo[2]{#2}
\providecommand\natexlab[1]{#1}
\providecommand\showeprint[2][]{arXiv:#2}

\bibitem[Akbari et~al\mbox{.}(2021)]%
        {akbari2021vatt}
\bibfield{author}{\bibinfo{person}{Hassan Akbari}, \bibinfo{person}{Liangzhe
  Yuan}, \bibinfo{person}{Rui Qian}, \bibinfo{person}{Wei-Hong Chuang},
  \bibinfo{person}{Shih-Fu Chang}, \bibinfo{person}{Yin Cui}, {and}
  \bibinfo{person}{Boqing Gong}.} \bibinfo{year}{2021}\natexlab{}.
\newblock \showarticletitle{Vatt: Transformers for multimodal self-supervised
  learning from raw video, audio and text}. In \bibinfo{booktitle}{\emph{Proc.
  NIPS}}, Vol.~\bibinfo{volume}{34}. \bibinfo{pages}{24206--24221}.
\newblock


\bibitem[Al-Tahan and Mohsenzadeh(2021)]%
        {al2021clar}
\bibfield{author}{\bibinfo{person}{Haider Al-Tahan} {and}
  \bibinfo{person}{Yalda Mohsenzadeh}.} \bibinfo{year}{2021}\natexlab{}.
\newblock \showarticletitle{Clar: Contrastive learning of auditory
  representations}. In \bibinfo{booktitle}{\emph{Proc. AISTATS}}. PMLR,
  \bibinfo{pages}{2530--2538}.
\newblock


\bibitem[Alayrac et~al\mbox{.}(2020)]%
        {alayrac2020self}
\bibfield{author}{\bibinfo{person}{Jean-Baptiste Alayrac},
  \bibinfo{person}{Adria Recasens}, \bibinfo{person}{Rosalia Schneider},
  \bibinfo{person}{Relja Arandjelovi{\'c}}, \bibinfo{person}{Jason Ramapuram},
  \bibinfo{person}{Jeffrey De~Fauw}, \bibinfo{person}{Lucas Smaira},
  \bibinfo{person}{Sander Dieleman}, {and} \bibinfo{person}{Andrew Zisserman}.}
  \bibinfo{year}{2020}\natexlab{}.
\newblock \showarticletitle{Self-supervised multimodal versatile networks}. In
  \bibinfo{booktitle}{\emph{Proc. NIPS}}, Vol.~\bibinfo{volume}{33}.
  \bibinfo{pages}{25--37}.
\newblock


\bibitem[Antol et~al\mbox{.}(2015)]%
        {antol2015vqa}
\bibfield{author}{\bibinfo{person}{Stanislaw Antol}, \bibinfo{person}{Aishwarya
  Agrawal}, \bibinfo{person}{Jiasen Lu}, \bibinfo{person}{Margaret Mitchell},
  \bibinfo{person}{Dhruv Batra}, \bibinfo{person}{C~Lawrence Zitnick}, {and}
  \bibinfo{person}{Devi Parikh}.} \bibinfo{year}{2015}\natexlab{}.
\newblock \showarticletitle{Vqa: Visual question answering}. In
  \bibinfo{booktitle}{\emph{Proc. ICCV}}. \bibinfo{pages}{2425--2433}.
\newblock


\bibitem[Baevski et~al\mbox{.}(2020)]%
        {baevski2020wav2vec}
\bibfield{author}{\bibinfo{person}{Alexei Baevski}, \bibinfo{person}{Yuhao
  Zhou}, \bibinfo{person}{Abdelrahman Mohamed}, {and} \bibinfo{person}{Michael
  Auli}.} \bibinfo{year}{2020}\natexlab{}.
\newblock \showarticletitle{wav2vec 2.0: A framework for self-supervised
  learning of speech representations}. In \bibinfo{booktitle}{\emph{Proc.
  NIPS}}, Vol.~\bibinfo{volume}{33}. \bibinfo{pages}{12449--12460}.
\newblock


\bibitem[Cakir et~al\mbox{.}(2015)]%
        {cakir2015polyphonic}
\bibfield{author}{\bibinfo{person}{Emre Cakir}, \bibinfo{person}{Toni
  Heittola}, \bibinfo{person}{Heikki Huttunen}, {and} \bibinfo{person}{Tuomas
  Virtanen}.} \bibinfo{year}{2015}\natexlab{}.
\newblock \showarticletitle{Polyphonic sound event detection using multi label
  deep neural networks}. In \bibinfo{booktitle}{\emph{Proc. IJCNN}}.
  \bibinfo{pages}{1--7}.
\newblock


\bibitem[Chen et~al\mbox{.}(2020c)]%
        {chen2020vggsound}
\bibfield{author}{\bibinfo{person}{Honglie Chen}, \bibinfo{person}{Weidi Xie},
  \bibinfo{person}{Andrea Vedaldi}, {and} \bibinfo{person}{Andrew Zisserman}.}
  \bibinfo{year}{2020}\natexlab{c}.
\newblock \showarticletitle{Vggsound: A large-scale audio-visual dataset}. In
  \bibinfo{booktitle}{\emph{Proc. IEEE ICASSP}}. \bibinfo{pages}{721--725}.
\newblock


\bibitem[Chen et~al\mbox{.}(2020b)]%
        {chen2020audio}
\bibfield{author}{\bibinfo{person}{Kun Chen}, \bibinfo{person}{Yusong Wu},
  \bibinfo{person}{Ziyue Wang}, \bibinfo{person}{Xuan Zhang},
  \bibinfo{person}{Fudong Nian}, \bibinfo{person}{Shengchen Li}, {and}
  \bibinfo{person}{Xi Shao}.} \bibinfo{year}{2020}\natexlab{b}.
\newblock \showarticletitle{Audio Captioning Based on Transformer and
  Pre-Trained CNN.}. In \bibinfo{booktitle}{\emph{Proc. DCASE}}.
  \bibinfo{pages}{21--25}.
\newblock


\bibitem[Chen et~al\mbox{.}(2022)]%
        {chen2022wavlm}
\bibfield{author}{\bibinfo{person}{Sanyuan Chen}, \bibinfo{person}{Chengyi
  Wang}, \bibinfo{person}{Zhengyang Chen}, \bibinfo{person}{Yu Wu},
  \bibinfo{person}{Shujie Liu}, \bibinfo{person}{Zhuo Chen},
  \bibinfo{person}{Jinyu Li}, \bibinfo{person}{Naoyuki Kanda},
  \bibinfo{person}{Takuya Yoshioka}, \bibinfo{person}{Xiong Xiao},
  {et~al\mbox{.}}} \bibinfo{year}{2022}\natexlab{}.
\newblock \showarticletitle{Wavlm: Large-scale self-supervised pre-training for
  full stack speech processing}.
\newblock \bibinfo{journal}{\emph{IEEE JSTSP}} \bibinfo{volume}{16},
  \bibinfo{number}{6} (\bibinfo{year}{2022}), \bibinfo{pages}{1505--1518}.
\newblock


\bibitem[Chen et~al\mbox{.}(2020a)]%
        {chen2020uniter}
\bibfield{author}{\bibinfo{person}{Yen-Chun Chen}, \bibinfo{person}{Linjie Li},
  \bibinfo{person}{Licheng Yu}, \bibinfo{person}{Ahmed El~Kholy},
  \bibinfo{person}{Faisal Ahmed}, \bibinfo{person}{Zhe Gan},
  \bibinfo{person}{Yu Cheng}, {and} \bibinfo{person}{Jingjing Liu}.}
  \bibinfo{year}{2020}\natexlab{a}.
\newblock \showarticletitle{Uniter: Universal image-text representation
  learning}. In \bibinfo{booktitle}{\emph{Proc. ECCV}}. Springer,
  \bibinfo{pages}{104--120}.
\newblock


\bibitem[den Oord et~al\mbox{.}(2018)]%
        {van2018representation}
\bibfield{author}{\bibinfo{person}{Van den Oord} {et~al\mbox{.}}}
  \bibinfo{year}{2018}\natexlab{}.
\newblock \showarticletitle{Representation learning with contrastive predictive
  coding}.
\newblock \bibinfo{journal}{\emph{arXiv preprint arXiv:1807.03748}}
  \bibinfo{volume}{2}, \bibinfo{number}{3} (\bibinfo{year}{2018}),
  \bibinfo{pages}{4}.
\newblock


\bibitem[Devlin et~al\mbox{.}(2019)]%
        {devlin2019bert}
\bibfield{author}{\bibinfo{person}{Jacob Devlin}, \bibinfo{person}{Ming-Wei
  Chang}, \bibinfo{person}{Kenton Lee}, {and} \bibinfo{person}{Kristin
  Toutanova}.} \bibinfo{year}{2019}\natexlab{}.
\newblock \showarticletitle{BERT: Pre-training of Deep Bidirectional
  Transformers for Language Understanding}. In \bibinfo{booktitle}{\emph{Proc.
  NAACL}}. \bibinfo{pages}{4171--4186}.
\newblock


\bibitem[Drossos et~al\mbox{.}(2017)]%
        {drossos2017automated}
\bibfield{author}{\bibinfo{person}{Konstantinos Drossos},
  \bibinfo{person}{Sharath Adavanne}, {and} \bibinfo{person}{Tuomas Virtanen}.}
  \bibinfo{year}{2017}\natexlab{}.
\newblock \showarticletitle{Automated audio captioning with recurrent neural
  networks}. In \bibinfo{booktitle}{\emph{Proc. IEEE WASPAA}}.
  \bibinfo{pages}{374--378}.
\newblock


\bibitem[Drossos et~al\mbox{.}(2020)]%
        {drossos2020clotho}
\bibfield{author}{\bibinfo{person}{Konstantinos Drossos},
  \bibinfo{person}{Samuel Lipping}, {and} \bibinfo{person}{Tuomas Virtanen}.}
  \bibinfo{year}{2020}\natexlab{}.
\newblock \showarticletitle{Clotho: An audio captioning dataset}. In
  \bibinfo{booktitle}{\emph{Proc. IEEE ICASSP}}. \bibinfo{pages}{736--740}.
\newblock


\bibitem[Elizalde et~al\mbox{.}(2023)]%
        {elizalde2023clap}
\bibfield{author}{\bibinfo{person}{Benjamin Elizalde}, \bibinfo{person}{Soham
  Deshmukh}, \bibinfo{person}{Mahmoud Al~Ismail}, {and}
  \bibinfo{person}{Huaming Wang}.} \bibinfo{year}{2023}\natexlab{}.
\newblock \showarticletitle{Clap learning audio concepts from natural language
  supervision}. In \bibinfo{booktitle}{\emph{Proc. IEEE ICASSP}}. IEEE,
  \bibinfo{pages}{1--5}.
\newblock


\bibitem[Eren and Sert(2020)]%
        {eren2020semantic}
\bibfield{author}{\bibinfo{person}{Ay{\c{s}}eg{\"u}l~{\"O}zkaya Eren} {and}
  \bibinfo{person}{Mustafa Sert}.} \bibinfo{year}{2020}\natexlab{}.
\newblock \showarticletitle{{Audio Captioning Based on Combined Audio and
  Semantic Embeddings}}. In \bibinfo{booktitle}{\emph{Proc. ISM}}.
  \bibinfo{pages}{41--48}.
\newblock


\bibitem[Fonseca et~al\mbox{.}(2022)]%
        {fonseca2022fsd50k}
\bibfield{author}{\bibinfo{person}{Eduardo Fonseca}, \bibinfo{person}{Xavier
  Favory}, \bibinfo{person}{Jordi Pons}, \bibinfo{person}{Frederic Font}, {and}
  \bibinfo{person}{Xavier Serra}.} \bibinfo{year}{2022}\natexlab{}.
\newblock \showarticletitle{Fsd50k: an open dataset of human-labeled sound
  events}.
\newblock \bibinfo{journal}{\emph{IEEE/ACM TASLP}}  \bibinfo{volume}{30}
  (\bibinfo{year}{2022}), \bibinfo{pages}{829--852}.
\newblock


\bibitem[Font et~al\mbox{.}(2013)]%
        {font2013freesound}
\bibfield{author}{\bibinfo{person}{Frederic Font}, \bibinfo{person}{Gerard
  Roma}, {and} \bibinfo{person}{Xavier Serra}.}
  \bibinfo{year}{2013}\natexlab{}.
\newblock \showarticletitle{Freesound technical demo}. In
  \bibinfo{booktitle}{\emph{Proc. ACM MM}}. \bibinfo{pages}{411--412}.
\newblock


\bibitem[Gemmeke et~al\mbox{.}(2017)]%
        {gemmeke2017audio}
\bibfield{author}{\bibinfo{person}{Jort~F Gemmeke}, \bibinfo{person}{Daniel~PW
  Ellis}, \bibinfo{person}{Dylan Freedman}, \bibinfo{person}{Aren Jansen},
  \bibinfo{person}{Wade Lawrence}, \bibinfo{person}{R~Channing Moore},
  \bibinfo{person}{Manoj Plakal}, {and} \bibinfo{person}{Marvin Ritter}.}
  \bibinfo{year}{2017}\natexlab{}.
\newblock \showarticletitle{Audio set: An ontology and human-labeled dataset
  for audio events}. In \bibinfo{booktitle}{\emph{Proc. IEEE ICASSP}}.
  \bibinfo{pages}{776--780}.
\newblock


\bibitem[Gong et~al\mbox{.}(2021a)]%
        {gong2021audio}
\bibfield{author}{\bibinfo{person}{Yuan Gong}, \bibinfo{person}{Yu~An Chung},
  {and} \bibinfo{person}{James Glass}.} \bibinfo{year}{2021}\natexlab{a}.
\newblock \showarticletitle{AST: Audio Spectrogram Transformer}. In
  \bibinfo{booktitle}{\emph{Proc. ISCA Interspeech}}. \bibinfo{pages}{56--60}.
\newblock


\bibitem[Gong et~al\mbox{.}(2021b)]%
        {gong2021psla}
\bibfield{author}{\bibinfo{person}{Yuan Gong}, \bibinfo{person}{Yu-An Chung},
  {and} \bibinfo{person}{James Glass}.} \bibinfo{year}{2021}\natexlab{b}.
\newblock \showarticletitle{{PSLA: Improving Audio Tagging With Pretraining,
  Sampling, Labeling, and Aggregation}}.
\newblock \bibinfo{journal}{\emph{IEEE/ACM TASLP}}  \bibinfo{volume}{29}
  (\bibinfo{year}{2021}), \bibinfo{pages}{3292--3306}.
\newblock


\bibitem[Guzhov et~al\mbox{.}(2022)]%
        {guzhov2022audioclip}
\bibfield{author}{\bibinfo{person}{Andrey Guzhov}, \bibinfo{person}{Federico
  Raue}, \bibinfo{person}{J{\"o}rn Hees}, {and} \bibinfo{person}{Andreas
  Dengel}.} \bibinfo{year}{2022}\natexlab{}.
\newblock \showarticletitle{Audioclip: Extending clip to image, text and
  audio}. In \bibinfo{booktitle}{\emph{Proc. IEEE ICASSP}}.
  \bibinfo{pages}{976--980}.
\newblock


\bibitem[Hsu et~al\mbox{.}(2021)]%
        {hsu2021hubert}
\bibfield{author}{\bibinfo{person}{Wei-Ning Hsu}, \bibinfo{person}{Benjamin
  Bolte}, \bibinfo{person}{Yao-Hung~Hubert Tsai}, \bibinfo{person}{Kushal
  Lakhotia}, \bibinfo{person}{Ruslan Salakhutdinov}, {and}
  \bibinfo{person}{Abdelrahman Mohamed}.} \bibinfo{year}{2021}\natexlab{}.
\newblock \showarticletitle{Hubert: Self-supervised speech representation
  learning by masked prediction of hidden units}.
\newblock \bibinfo{journal}{\emph{IEEE/ACM TASLP}}  \bibinfo{volume}{29}
  (\bibinfo{year}{2021}), \bibinfo{pages}{3451--3460}.
\newblock


\bibitem[Jia et~al\mbox{.}(2021)]%
        {jia2021scaling}
\bibfield{author}{\bibinfo{person}{Chao Jia}, \bibinfo{person}{Yinfei Yang},
  \bibinfo{person}{Ye Xia}, \bibinfo{person}{Yi-Ting Chen},
  \bibinfo{person}{Zarana Parekh}, \bibinfo{person}{Hieu Pham},
  \bibinfo{person}{Quoc Le}, \bibinfo{person}{Yun-Hsuan Sung},
  \bibinfo{person}{Zhen Li}, {and} \bibinfo{person}{Tom Duerig}.}
  \bibinfo{year}{2021}\natexlab{}.
\newblock \showarticletitle{Scaling up visual and vision-language
  representation learning with noisy text supervision}. In
  \bibinfo{booktitle}{\emph{Proc. ICML}}. PMLR, \bibinfo{pages}{4904--4916}.
\newblock


\bibitem[Kazakos et~al\mbox{.}(2021)]%
        {kazakos2021slow}
\bibfield{author}{\bibinfo{person}{Evangelos Kazakos}, \bibinfo{person}{Arsha
  Nagrani}, \bibinfo{person}{Andrew Zisserman}, {and} \bibinfo{person}{Dima
  Damen}.} \bibinfo{year}{2021}\natexlab{}.
\newblock \showarticletitle{Slow-fast auditory streams for audio recognition}.
  In \bibinfo{booktitle}{\emph{Proc. IEEE ICASSP}}. \bibinfo{pages}{855--859}.
\newblock


\bibitem[Kim et~al\mbox{.}(2019)]%
        {kim2019audiocaps}
\bibfield{author}{\bibinfo{person}{Chris~Dongjoo Kim},
  \bibinfo{person}{Byeongchang Kim}, \bibinfo{person}{Hyunmin Lee}, {and}
  \bibinfo{person}{Gunhee Kim}.} \bibinfo{year}{2019}\natexlab{}.
\newblock \showarticletitle{AudioCaps: Generating Captions for Audios in The
  Wild}. In \bibinfo{booktitle}{\emph{Proc. NAACL}}. \bibinfo{pages}{119--132}.
\newblock


\bibitem[Koizumi et~al\mbox{.}(2020)]%
        {koizumi2020audio}
\bibfield{author}{\bibinfo{person}{Yuma Koizumi}, \bibinfo{person}{Yasunori
  Ohishi}, \bibinfo{person}{Daisuke Niizumi}, \bibinfo{person}{Daiki Takeuchi},
  {and} \bibinfo{person}{Masahiro Yasuda}.} \bibinfo{year}{2020}\natexlab{}.
\newblock \showarticletitle{Audio Captioning using Pre-Trained Large-Scale
  Language Model Guided by Audio-based Similar Caption Retrieval}.
\newblock \bibinfo{journal}{\emph{arXiv preprint arXiv:2012.07331}}
  (\bibinfo{year}{2020}).
\newblock


\bibitem[Kong et~al\mbox{.}(2020)]%
        {kong2020panns}
\bibfield{author}{\bibinfo{person}{Qiuqiang Kong}, \bibinfo{person}{Yin Cao},
  \bibinfo{person}{Turab Iqbal}, \bibinfo{person}{Yuxuan Wang},
  \bibinfo{person}{Wenwu Wang}, {and} \bibinfo{person}{Mark~D Plumbley}.}
  \bibinfo{year}{2020}\natexlab{}.
\newblock \showarticletitle{Panns: Large-scale pretrained audio neural networks
  for audio pattern recognition}.
\newblock \bibinfo{journal}{\emph{IEEE/ACM TASLP}}  \bibinfo{volume}{28}
  (\bibinfo{year}{2020}), \bibinfo{pages}{2880--2894}.
\newblock


\bibitem[Koutini et~al\mbox{.}(2022)]%
        {koutini2022efficient}
\bibfield{author}{\bibinfo{person}{Khaled Koutini}, \bibinfo{person}{Jan
  Schl{\"u}ter}, \bibinfo{person}{Hamid Eghbal-zadeh}, {and}
  \bibinfo{person}{Gerhard Widmer}.} \bibinfo{year}{2022}\natexlab{}.
\newblock \showarticletitle{Efficient Training of Audio Transformers with
  Patchout}. In \bibinfo{booktitle}{\emph{Proc. ISCA Interspeech}}.
  \bibinfo{pages}{2753--2757}.
\newblock


\bibitem[Krishna et~al\mbox{.}(2017)]%
        {krishna2017visual}
\bibfield{author}{\bibinfo{person}{Ranjay Krishna}, \bibinfo{person}{Yuke Zhu},
  \bibinfo{person}{Oliver Groth}, \bibinfo{person}{Justin Johnson},
  \bibinfo{person}{Kenji Hata}, \bibinfo{person}{Joshua Kravitz},
  \bibinfo{person}{Stephanie Chen}, \bibinfo{person}{Yannis Kalantidis},
  \bibinfo{person}{Li-Jia Li}, \bibinfo{person}{David~A Shamma},
  {et~al\mbox{.}}} \bibinfo{year}{2017}\natexlab{}.
\newblock \showarticletitle{Visual genome: Connecting language and vision using
  crowdsourced dense image annotations}.
\newblock \bibinfo{journal}{\emph{IJCV}} \bibinfo{volume}{123},
  \bibinfo{number}{1} (\bibinfo{year}{2017}), \bibinfo{pages}{32--73}.
\newblock


\bibitem[Li et~al\mbox{.}(2020)]%
        {li2020oscar}
\bibfield{author}{\bibinfo{person}{Xiujun Li}, \bibinfo{person}{Xi Yin},
  \bibinfo{person}{Chunyuan Li}, \bibinfo{person}{Pengchuan Zhang},
  \bibinfo{person}{Xiaowei Hu}, \bibinfo{person}{Lei Zhang},
  \bibinfo{person}{Lijuan Wang}, \bibinfo{person}{Houdong Hu},
  \bibinfo{person}{Li Dong}, \bibinfo{person}{Furu Wei}, {et~al\mbox{.}}}
  \bibinfo{year}{2020}\natexlab{}.
\newblock \showarticletitle{Oscar: Object-semantics aligned pre-training for
  vision-language tasks}. In \bibinfo{booktitle}{\emph{Proc. ECCV}}. Springer,
  \bibinfo{pages}{121--137}.
\newblock


\bibitem[Lin et~al\mbox{.}(2014)]%
        {lin2014microsoft}
\bibfield{author}{\bibinfo{person}{Tsung-Yi Lin}, \bibinfo{person}{Michael
  Maire}, \bibinfo{person}{Serge Belongie}, \bibinfo{person}{James Hays},
  \bibinfo{person}{Pietro Perona}, \bibinfo{person}{Deva Ramanan},
  \bibinfo{person}{Piotr Doll{\'a}r}, {and} \bibinfo{person}{C~Lawrence
  Zitnick}.} \bibinfo{year}{2014}\natexlab{}.
\newblock \showarticletitle{Microsoft coco: Common objects in context}. In
  \bibinfo{booktitle}{\emph{Proc. ECCV}}. Springer, \bibinfo{pages}{740--755}.
\newblock


\bibitem[Loshchilov and Hutter(2016)]%
        {loshchilov2016sgdr}
\bibfield{author}{\bibinfo{person}{Ilya Loshchilov} {and}
  \bibinfo{person}{Frank Hutter}.} \bibinfo{year}{2016}\natexlab{}.
\newblock \showarticletitle{Sgdr: Stochastic gradient descent with warm
  restarts}.
\newblock \bibinfo{journal}{\emph{arXiv preprint arXiv:1608.03983}}
  (\bibinfo{year}{2016}).
\newblock


\bibitem[Lu et~al\mbox{.}(2019)]%
        {lu2019vilbert}
\bibfield{author}{\bibinfo{person}{Jiasen Lu}, \bibinfo{person}{Dhruv Batra},
  \bibinfo{person}{Devi Parikh}, {and} \bibinfo{person}{Stefan Lee}.}
  \bibinfo{year}{2019}\natexlab{}.
\newblock \showarticletitle{ViLBERT: pretraining task-agnostic visiolinguistic
  representations for vision-and-language tasks}. In
  \bibinfo{booktitle}{\emph{Proc. NIPS}}. \bibinfo{pages}{13--23}.
\newblock


\bibitem[Martin and Mesaros(2021)]%
        {martin2021diversity}
\bibfield{author}{\bibinfo{person}{Irene Martin} {and}
  \bibinfo{person}{Annamaria Mesaros}.} \bibinfo{year}{2021}\natexlab{}.
\newblock \showarticletitle{Diversity and Bias in Audio Captioning Datasets}.
  In \bibinfo{booktitle}{\emph{Proc. DCASE}}. \bibinfo{pages}{90--94}.
\newblock


\bibitem[Mesaros et~al\mbox{.}(2018)]%
        {mesaros2018multi}
\bibfield{author}{\bibinfo{person}{Annamaria Mesaros}, \bibinfo{person}{Toni
  Heittola}, {and} \bibinfo{person}{Tuomas Virtanen}.}
  \bibinfo{year}{2018}\natexlab{}.
\newblock \showarticletitle{A multi-device dataset for urban acoustic scene
  classification}. In \bibinfo{booktitle}{\emph{Proc. DCASE}}.
  \bibinfo{pages}{9--13}.
\newblock


\bibitem[Miech et~al\mbox{.}(2019)]%
        {miech2019howto100m}
\bibfield{author}{\bibinfo{person}{Antoine Miech}, \bibinfo{person}{Dimitri
  Zhukov}, \bibinfo{person}{Jean-Baptiste Alayrac}, \bibinfo{person}{Makarand
  Tapaswi}, \bibinfo{person}{Ivan Laptev}, {and} \bibinfo{person}{Josef
  Sivic}.} \bibinfo{year}{2019}\natexlab{}.
\newblock \showarticletitle{Howto100m: Learning a text-video embedding by
  watching hundred million narrated video clips}. In
  \bibinfo{booktitle}{\emph{Proc. ICCV}}. \bibinfo{pages}{2630--2640}.
\newblock


\bibitem[Niizumi et~al\mbox{.}(2021)]%
        {niizumi2021byol}
\bibfield{author}{\bibinfo{person}{Daisuke Niizumi}, \bibinfo{person}{Daiki
  Takeuchi}, \bibinfo{person}{Yasunori Ohishi}, \bibinfo{person}{Noboru
  Harada}, {and} \bibinfo{person}{Kunio Kashino}.}
  \bibinfo{year}{2021}\natexlab{}.
\newblock \showarticletitle{BYOL for audio: Self-supervised learning for
  general-purpose audio representation}. In \bibinfo{booktitle}{\emph{Proc.
  IJCNN}}. \bibinfo{pages}{1--8}.
\newblock


\bibitem[Oncescu et~al\mbox{.}(2021)]%
        {oncescu21audio}
\bibfield{author}{\bibinfo{person}{Andreea-Maria Oncescu},
  \bibinfo{person}{A.~Sophia Koepke}, \bibinfo{person}{João~F. Henriques},
  \bibinfo{person}{Zeynep Akata}, {and} \bibinfo{person}{Samuel Albanie}.}
  \bibinfo{year}{2021}\natexlab{}.
\newblock \showarticletitle{{Audio Retrieval with Natural Language Queries}}.
  In \bibinfo{booktitle}{\emph{Proc. ISCA Interspeech}}.
  \bibinfo{pages}{2411--2415}.
\newblock


\bibitem[Radford et~al\mbox{.}(2021)]%
        {radford2021learning}
\bibfield{author}{\bibinfo{person}{Alec Radford}, \bibinfo{person}{Jong~Wook
  Kim}, \bibinfo{person}{Chris Hallacy}, \bibinfo{person}{Aditya Ramesh},
  \bibinfo{person}{Gabriel Goh}, \bibinfo{person}{Sandhini Agarwal},
  \bibinfo{person}{Girish Sastry}, \bibinfo{person}{Amanda Askell},
  \bibinfo{person}{Pamela Mishkin}, \bibinfo{person}{Jack Clark},
  {et~al\mbox{.}}} \bibinfo{year}{2021}\natexlab{}.
\newblock \showarticletitle{Learning transferable visual models from natural
  language supervision}. In \bibinfo{booktitle}{\emph{Proc. ICML}}.
  \bibinfo{pages}{8748--8763}.
\newblock


\bibitem[Saeed et~al\mbox{.}(2021)]%
        {saeed2021contrastive}
\bibfield{author}{\bibinfo{person}{Aaqib Saeed}, \bibinfo{person}{David
  Grangier}, {and} \bibinfo{person}{Neil Zeghidour}.}
  \bibinfo{year}{2021}\natexlab{}.
\newblock \showarticletitle{Contrastive learning of general-purpose audio
  representations}. In \bibinfo{booktitle}{\emph{Proc. IEEE ICASSP}}.
  \bibinfo{pages}{3875--3879}.
\newblock


\bibitem[Sharma et~al\mbox{.}(2018)]%
        {sharma2018conceptual}
\bibfield{author}{\bibinfo{person}{Piyush Sharma}, \bibinfo{person}{Nan Ding},
  \bibinfo{person}{Sebastian Goodman}, {and} \bibinfo{person}{Radu Soricut}.}
  \bibinfo{year}{2018}\natexlab{}.
\newblock \showarticletitle{Conceptual captions: A cleaned, hypernymed, image
  alt-text dataset for automatic image captioning}. In
  \bibinfo{booktitle}{\emph{Proc. ACL}}. \bibinfo{pages}{2556--2565}.
\newblock


\bibitem[Su et~al\mbox{.}(2019)]%
        {su2019vl}
\bibfield{author}{\bibinfo{person}{Weijie Su}, \bibinfo{person}{Xizhou Zhu},
  \bibinfo{person}{Yue Cao}, \bibinfo{person}{Bin Li}, \bibinfo{person}{Lewei
  Lu}, \bibinfo{person}{Furu Wei}, {and} \bibinfo{person}{Jifeng Dai}.}
  \bibinfo{year}{2019}\natexlab{}.
\newblock \showarticletitle{VL-BERT: Pre-training of Generic Visual-Linguistic
  Representations}. In \bibinfo{booktitle}{\emph{Proc. ICLR}}.
  \bibinfo{pages}{1--16}.
\newblock


\bibitem[Turc et~al\mbox{.}(2019)]%
        {turc2019well}
\bibfield{author}{\bibinfo{person}{Iulia Turc}, \bibinfo{person}{Ming-Wei
  Chang}, \bibinfo{person}{Kenton Lee}, {and} \bibinfo{person}{Kristina
  Toutanova}.} \bibinfo{year}{2019}\natexlab{}.
\newblock \showarticletitle{Well-read students learn better: On the importance
  of pre-training compact models}.
\newblock \bibinfo{journal}{\emph{arXiv preprint arXiv:1908.08962}}
  (\bibinfo{year}{2019}).
\newblock


\bibitem[Wang et~al\mbox{.}(2022)]%
        {wang2022ofa}
\bibfield{author}{\bibinfo{person}{Peng Wang}, \bibinfo{person}{An Yang},
  \bibinfo{person}{Rui Men}, \bibinfo{person}{Junyang Lin},
  \bibinfo{person}{Shuai Bai}, \bibinfo{person}{Zhikang Li},
  \bibinfo{person}{Jianxin Ma}, \bibinfo{person}{Chang Zhou},
  \bibinfo{person}{Jingren Zhou}, {and} \bibinfo{person}{Hongxia Yang}.}
  \bibinfo{year}{2022}\natexlab{}.
\newblock \showarticletitle{Ofa: Unifying architectures, tasks, and modalities
  through a simple sequence-to-sequence learning framework}. In
  \bibinfo{booktitle}{\emph{Proc. ICML}}. PMLR, \bibinfo{pages}{23318--23340}.
\newblock


\bibitem[Wang et~al\mbox{.}(2021)]%
        {wang2021simvlm}
\bibfield{author}{\bibinfo{person}{Zirui Wang}, \bibinfo{person}{Jiahui Yu},
  \bibinfo{person}{Adams~Wei Yu}, \bibinfo{person}{Zihang Dai},
  \bibinfo{person}{Yulia Tsvetkov}, {and} \bibinfo{person}{Yuan Cao}.}
  \bibinfo{year}{2021}\natexlab{}.
\newblock \showarticletitle{SimVLM: Simple Visual Language Model Pretraining
  with Weak Supervision}. In \bibinfo{booktitle}{\emph{Proc. ICLR}}.
  \bibinfo{pages}{1--17}.
\newblock


\bibitem[Wu et~al\mbox{.}(2022)]%
        {wu2022wav2clip}
\bibfield{author}{\bibinfo{person}{Ho-Hsiang Wu}, \bibinfo{person}{Prem
  Seetharaman}, \bibinfo{person}{Kundan Kumar}, {and}
  \bibinfo{person}{Juan~Pablo Bello}.} \bibinfo{year}{2022}\natexlab{}.
\newblock \showarticletitle{Wav2CLIP: Learning Robust Audio Representations
  From CLIP}. In \bibinfo{booktitle}{\emph{Proc. IEEE ICASSP}}.
\newblock


\bibitem[Xu et~al\mbox{.}(2021)]%
        {xu2021investigating}
\bibfield{author}{\bibinfo{person}{Xuenan Xu}, \bibinfo{person}{Heinrich
  Dinkel}, \bibinfo{person}{Mengyue Wu}, \bibinfo{person}{Zeyu Xie}, {and}
  \bibinfo{person}{Kai Yu}.} \bibinfo{year}{2021}\natexlab{}.
\newblock \showarticletitle{Investigating local and global information for
  automated audio captioning with transfer learning}. In
  \bibinfo{booktitle}{\emph{Proc. IEEE ICASSP}}. \bibinfo{pages}{905--909}.
\newblock


\bibitem[Xu et~al\mbox{.}(2022)]%
        {xu2022sjtu}
\bibfield{author}{\bibinfo{person}{Xuenan Xu}, \bibinfo{person}{Zeyu Xie},
  \bibinfo{person}{Mengyue Wu}, {and} \bibinfo{person}{Kai Yu}.}
  \bibinfo{year}{2022}\natexlab{}.
\newblock \bibinfo{booktitle}{\emph{The {SJTU} System for {DCASE2022} Challenge
  Task 6: Audio Captioning with Audio-Text Retrieval Pre-training}}.
\newblock \bibinfo{type}{{T}echnical {R}eport}. \bibinfo{institution}{DCASE2022
  Challenge}.
\newblock


\bibitem[Zellers et~al\mbox{.}(2019)]%
        {zellers2019recognition}
\bibfield{author}{\bibinfo{person}{Rowan Zellers}, \bibinfo{person}{Yonatan
  Bisk}, \bibinfo{person}{Ali Farhadi}, {and} \bibinfo{person}{Yejin Choi}.}
  \bibinfo{year}{2019}\natexlab{}.
\newblock \showarticletitle{From recognition to cognition: Visual commonsense
  reasoning}. In \bibinfo{booktitle}{\emph{Proc. CVPR}}.
  \bibinfo{pages}{6720--6731}.
\newblock


\bibitem[Zhang et~al\mbox{.}(2021)]%
        {zhang2021enriching}
\bibfield{author}{\bibinfo{person}{Zhiling Zhang}, \bibinfo{person}{Zelin
  Zhou}, \bibinfo{person}{Haifeng Tang}, \bibinfo{person}{Guangwei Li},
  \bibinfo{person}{Mengyue Wu}, {and} \bibinfo{person}{Kenny~Q Zhu}.}
  \bibinfo{year}{2021}\natexlab{}.
\newblock \showarticletitle{Enriching Ontology with Temporal Commonsense for
  Low-Resource Audio Tagging}. In \bibinfo{booktitle}{\emph{Proc. CIKM}}.
  \bibinfo{pages}{3652--3656}.
\newblock


\bibitem[Zhao et~al\mbox{.}(2022)]%
        {zhao2022connecting}
\bibfield{author}{\bibinfo{person}{Yanpeng Zhao}, \bibinfo{person}{Jack
  Hessel}, \bibinfo{person}{Youngjae Yu}, \bibinfo{person}{Ximing Lu},
  \bibinfo{person}{Rowan Zellers}, {and} \bibinfo{person}{Yejin Choi}.}
  \bibinfo{year}{2022}\natexlab{}.
\newblock \showarticletitle{Connecting the Dots between Audio and Text without
  Parallel Data through Visual Knowledge Transfer}. In
  \bibinfo{booktitle}{\emph{Proc. NAACL}}. \bibinfo{pages}{4492--4507}.
\newblock


\bibitem[Zhou et~al\mbox{.}(2022)]%
        {zhou2022can}
\bibfield{author}{\bibinfo{person}{Zelin Zhou}, \bibinfo{person}{Zhiling
  Zhang}, \bibinfo{person}{Xuenan Xu}, \bibinfo{person}{Zeyu Xie},
  \bibinfo{person}{Mengyue Wu}, {and} \bibinfo{person}{Kenny~Q Zhu}.}
  \bibinfo{year}{2022}\natexlab{}.
\newblock \showarticletitle{Can audio captions be evaluated with image caption
  metrics?}. In \bibinfo{booktitle}{\emph{Proc. IEEE ICASSP}}.
  \bibinfo{pages}{981--985}.
\newblock


\end{thebibliography}

\end{document}